\documentclass{article}

\usepackage{amssymb,amsfonts,amsmath,stmaryrd}
\usepackage{cite,enumerate,float,indentfirst}
\usepackage{forloop}
\usepackage{stmaryrd}
\usepackage{color}
\usepackage{tikz}
\usetikzlibrary{arrows,snakes,backgrounds}
\usepackage{url}
\usepackage{hyperref}
\usepackage{textcomp}

\def\be{\begin{eqnarray}}
\def\ee{\end{eqnarray}}

\definecolor{red}{rgb}{1,0,0}
\definecolor{orange}{rgb}{1,0.5,0}
\definecolor{violet}{rgb}{0.7,0,1}

\hoffset= - 1.0in         

\newcount\hshift
\newcount\vshift
\newcount\tmp
\newcount\i
\newcount\j
\newcount\boxsize
\newcount\circleRadius
\newcount\xShift
\newcount\yShift
\newcount\len
\newcount\tmp
\newcount\prevElt
\newcount\curElt

\def\gridPut(#1,#2)#3{{
    \loccount\x
    \x=#1
    \multiply\x by \boxsize
    \loccount\y
    \y=#2
    \multiply\y by \boxsize
    \put(\x,\y){
      #3
    }
}}

\makeatletter
\newcommand\YD[1]{{
    \vshift=0
    \@for \elt:=#1 \do{
      \Yrow\vshift\elt
      \advance\vshift by -\boxsize
    }
}}
\makeatother

\def\Ybox{{
    \let\s\boxsize
    \put(\xShift, \yShift) {
      \put(0,0){\line(1,0){\s}}
      \put(\s,0){\line(0,1){\s}}
      \put(\s,\s){\line(-1,0){\s}}
      \put(0,\s){\line(0,-1){\s}}
    }
}}

\newcommand\Yrow[2]{{
    \hshift = 0
    \j = 0
    \loop \ifnum\j<#2
    \put(\hshift,#1){\Ybox}
    \advance\hshift by \boxsize
    \advance\j by 1
    \repeat
}}

\newcommand\Ycolumn[2]{{
    \vshift = 0
    \j = 0
    \loop \ifnum\j<#2
    \put(#1,\vshift){\Ybox}
    \advance\vshift by \boxsize
    \advance\j by 1
    \repeat
}}

\newcommand\youngEnv[1]{{
    \boxsize=10
    \circleRadius=5
    \xShift=0
    \yShift=0
    #1
}}

\makeatletter
\newcommand\inlineYD[1]{{
    \yShift=0
    \xShift=0
    \def\onlyFirst{\xShift=\elt \def\onlyFirst{}}
    \@for \elt:=#1 \do{
      \onlyFirst
      \advance\yShift by 1
    }
    \multiply\yShift by 5
    \multiply\xShift by 5
    \begin{picture}(\xShift,\yShift)(0,-\yShift)
      \youngEnv{\boxsize=5 \yShift=-5 \YD{#1}}
    \end{picture}
}}
\makeatother

\makeatletter
\newsavebox{\@brx}
\newcommand{\llangle}[1][]{\savebox{\@brx}{\(\m@th{#1\langle}\)}%
  \mathopen{\copy\@brx\kern-0.5\wd\@brx\usebox{\@brx}}}
\newcommand{\rrangle}[1][]{\savebox{\@brx}{\(\m@th{#1\rangle}\)}%
  \mathclose{\copy\@brx\kern-0.5\wd\@brx\usebox{\@brx}}}
\makeatother

\begin{document}

\title{\vspace{-1cm}{\Large {\bf
      On bilinear superintegrability for monomial matrix models in pure phase.
    }
    \date{}
}}

\maketitle
\vspace{-5.2cm}

\begin{center}
	\hfill ITEP/TH-25/23 \\
	\hfill IITP/TH-18/23 \\
	\hfill MIPT/TH-19/23
\end{center}

\vspace{2.2cm}

\begin{center}
  \begin{large}
    C.-T. Chan$^{a}$\footnote{ctchan@go.thu.edu.tw}
    V.Mishnyakov$^{b,c,d,e,f,g}$\footnote{mishnyakovvv@gmail.com},
    A. Popolitov$^{b,c,d,e}$\footnote{popolit@gmail.com},
    K. Tsybikov$^{b}$\footnote{tsybikov.kn@phystech.edu}
\end{large}
\end{center}

\begin{center}
  $^a$ {\small {\it Department of Applied Physics, Tunghai University, Taichung, 40704, Taiwan}}
  $^b$ {\small {\it Moscow Institute of Physics and Technology, Dolgoprudny 141701, Russia }} \\
  $^c$ {\small {\it NRC ``Kurchatov Institute''}}\\
  $^d$ {\small {\it Institute for Information Transmission Problems, Moscow 127994, Russia}}\\
  $^e$ {\small {\it ITEP, Moscow, Russia}} \\
  $^f$ {\small {\it Lebedev Physics Institute, Moscow 119991, Russia}} \\
  $^g$ {\small {\it Institute for Theoretical and Mathematical Physics, Lomonosov Moscow State University, Moscow 119991, Russia}} \\

\end{center}

\vspace{0.5cm}

\begin{abstract}
  We argue that the recently discovered bilinear superintegrability
  arXiv:2206.02045 generalizes, in a non-trivial way, to monomial matrix models
  in pure phase. The structure is much richer: for
  the trivial core Schur functions
  required modifications are minor, and the only new ingredient is a certain
  (contour-dependent) permutation matrix;
  for non-trivial-core Schur functions, in both bi-linear and tri-linear averages
  the deformation is more complicated: averages acquire extra N-dependent factors
  and selection rule is less straightforward to imply.
\end{abstract}

\bigskip

{\section{Introduction}\label{sec:introduction}
  We continue to implement the large program of concrete approach to
  quantum field theories. This program consists in the simple-to-complex study
  of ever complicating QFT setups, but each time in full generality with focus
  on non-perturbative phenomena and finite
  (neither infinitesimal not infinite) coupling constants regime.
  The hope is that arising essential complications are this way untangled
  and can be dealt with one by one.

  Our main focus is the bilinear superintegrability structure
  \cite{paper:MM-bilinear-character-correlators-in-superintegrable-theory}
  -- a generalization
  of usual, linear, superintegrability. The linear superintegrability itself was
  recently realized to be convenient language of non-perturbative, finite $N$, description
  of wide range of matrix models, in different regimes (phases)
  \cite{paper:CHPS-orbifolds-and-exact-solutions}.
  And the bilinear superintegrability, perhaps, even more importantly, sheds light
  on the previously obscure origins of the celebrated Nekrasov calculus
  \cite{paper:MM-superintegrability-as-the-hidden-origin-of-the-nekrasov-calculus}:
  the most fruitful concrete approach to non-perturbative physics of
  supersymmetric gauge theories \cite{paper:FJWZ-the-art-of-pestunization}.

  Specifically, we explain that bilinear superintegrability is not restricted to just
  Gaussian and logarithmic (Penner-like) models, but instead is more universal and,
  in particular, straightforwardly generalizes to the wide class of monomial matrix models
  in pure phase \cite{paper:CHPS-orbifolds-and-exact-solutions}.
  This is a wide class of models indeed, as any polynomial
  model observable can be expanded near suitable monomial point in \textit{convergent}
  power series; as opposed to usual asymptotic power series of perturbation theory
  near Gaussian (quadratic) point. The main statements
  are presented in Section~\ref{sec:main-statements}. The central role is played
  by the relevant monomial deformation of the box-factor-inserting operator $\mathcal{O}$
  (see \eqref{eq:calo-op-def}), which gradually seems to become one of the key objects
  in modern MM framework
  \cite{paper:MMMPWZ-interpolating-matrix-models-for-wlzz-series,
    paper:MMMPZ-on-kp-integrable-skew-hurwitz-tau-functions-and-their-beta-deformations,
    paper:LMMMPWZ-qt-deformed-skew-hurwitz-tau-functions,
    paper:MMMP-commutative-families-in-w-infty,
    paper:MMMP-commutative-subalgebras-for-serre-relations},
  which is being developed as the adequate language for understanding
  the recently proposed WLZZ models
  \cite{paper:WZZZ-cft-approach-to-constraint-operators-for-beta-deformed-homms,
    paper:WLZZ-superintegrability-for-beta-deformed-partition-function-hierarchies-with-w-representations} and their various natural generalizations.
  These concrete observations about the structure of bilinear superintegrable averages
  in monomial matrix models' pure phase constitute \textbf{the main result}
  of the present paper.

  The bilinear superintegrability most famously appears in (generalized)
  Kadell integrals (see~eqns.(5.1)--(5.3)~in~\cite{paper:CHPS-orbifolds-and-exact-solutions}),
  where the bilinear average of two Schur functions, one of them of shifted argument,
  in Dotsenko-Fateev (DF) type logarithmic model is equal to manifest factorized expression.
  
  \noindent However, the de-log~(~$v\rightarrow \infty$, $\log(1 - v X^r) \sim - X^r$~) limit
  of this formula, which restores the usual monomial potential, destroys the bilinearity
  of the correlator -- the shift becomes infinite. So, \textit{naively},
  the bilinear superintegrability formula in non-logarithmic monomial matrix models
  does not exist. However, if one believes that structures persist when taking simplification limits (and de-log \textit{is} a certain simplification) then bilinear superintegrability
  \textit{should} exist in this case.

  From this point-of-view, our formula
  \eqref{eq:bilinear-superintegrability} is the long awaited answer to this apparent
  puzzle: in the limit the ``shifted'' Schur polynomial becomes the ``associated''
  $K_\Delta$ polynomial (whose explicit formula \eqref{eq:k-delta-def}
  features a kind of shift operator
  in time-variables) and non-trivial (anomaly-like) permutation operation
  $\pi(\Delta)$ appears.

  Further, the simple form of single- and double- $K_\Delta$ averages
  \eqref{eq:single-k-average} and \eqref{eq:double-k-average}
  is reminiscent of the structure of the CFT correlators. Therefore,
  in Section~\ref{sec:triple-k-averages} we study the structure of triple-$K$ averages.
  It turns out to be more complicated than the naive expectation from CFT analogy,
  so the naive motto
  \begin{align}
    \text{Monomial MM in $K_\Delta$ basis} \equiv \text{some CFT}
  \end{align}
  is \textbf{wrong}. Still, the appearing non-factorizability seems tame enough
  (at most quadratic factors appear in studied examples) to deserve further intensive
  investigation.

  Finally, in Section~\ref{sec:proof-attempts} we summarize our proof attempts.
  It turns out that, while the single-average formula \eqref{eq:single-k-average}
  and implication
  $\eqref{eq:bilinear-superintegrability} \rightarrow \eqref{eq:double-k-average}$
  are quite straightforward, equally concise explanation for
  \eqref{eq:bilinear-superintegrability} itself is so far missing.
  This, of course, makes the existence of \eqref{eq:bilinear-superintegrability}
  even more valuable and non-trivial.
  
  \bigskip
  
  In this paper, as becomes customary for the papers about monomial matrix models,
  we freely use the language related to quotient division
  of partition by an integer $r$: $r$-cores, $r$-quotients, $r$-signatures,
  rim-hooks and so on.
  We refer the reader to Appendix A of \cite{paper:CHPS-orbifolds-and-exact-solutions},
  as well as to the original Macdonald book
  \cite{book:M-symmetric-functions-and-hall-polynomials}.
  
}

{\section{Main statements} \label{sec:main-statements}
  Monomial matrix model in pure phase can be \textit{defined} directly through
  its normalized Schur polynomial average
  \begin{align} \label{eq:monomial-schur-average}
    \left\langle S_R \right\rangle = & \ S_R \{\delta_{k,r}\} \frac{1}{r^{|R|/r}}
    \Lambda_{r,a}^R (N),
  \end{align}
  where $S_R \{\delta_{k,r}\}$ is the Schur polynomial evaluated at special point
  $p_k = \delta_{k,r}$,
  
  \noindent $\Lambda_{r,a}^R (N)$ is a peculiar product over boxes of the diagram $R$
  \begin{align} \label{eq:box-prod-def}
    \Lambda_{r,a}^R (N) = & \ \prod_{(i,j)\in R}
    \left [\left [ N-i+j \right]\right]_{r,0}
    \left [\left [ N-i+j \right]\right]_{r,a}, \text{ with } \\ \notag
    \left [\left [ f(i,j) \right]\right]_{r,x}
    = & \ f(i,j) \text{ if } f(i,j)-x \text{ mod } r = 0 \text{ else } 0,
  \end{align}
  that will frequently reappear in our presentation;
  $r$ is an integer $\geq 2$ and parameter $a$ runs from $0$ to $r-1$.
  The emergent additional parameter $b = N \text{ mod } r$ can be equal to $0$ or $a$.
  \footnote{The case of generic $b$ (the exotic pure phase, see
  \cite{paper:BP-null-a-superintegrability}) is, for simplicity,
  left out of the present text's scope and deserves separate study}
  
  \noindent Indeed, given \eqref{eq:monomial-schur-average}, normalized correlator of any other
  symmetric polynomial can be calculated as a linear combination of these, basis, ones.

  Motivated by the numerous papers on WLZZ models
  \cite{paper:MMMPWZ-interpolating-matrix-models-for-wlzz-series,
    paper:MMMPZ-on-kp-integrable-skew-hurwitz-tau-functions-and-their-beta-deformations,
    paper:LMMMPWZ-qt-deformed-skew-hurwitz-tau-functions,
    paper:MMMP-commutative-families-in-w-infty,
    paper:MMMP-commutative-subalgebras-for-serre-relations},
  we also frequently use the shorthand notation
  \begin{align}
    \xi_\Delta := \Lambda_{r,a}^R (N),
  \end{align}
  keeping in mind that in our case the $\xi$-factor depends on $N$, $r$, $a$ (and $b$).
  
  For the relation to the usual matrix model definition,
  through repeated integration see
  \cite{paper:CHPS-orbifolds-and-exact-solutions}
  and a more recent development \cite{paper:BP-null-a-superintegrability}.

  \bigskip

  Now consider auxiliary (associated) polynomials $K_\Delta$, which
  are related to Schur polynomials by manifest triangular change of variables
  \begin{align} \label{eq:k-delta-def}
    K_\Delta = \mathcal{O}^{-1}
    \exp\left((- r) \frac{\partial}{\partial p_r}\right)
    \mathcal{O} S_\Delta,
  \end{align}
  where $\mathcal{O}$-operator (resp. $\mathcal{O}^{-1}$-operator)
  is the operator that multiplies (resp. divides) each Schur function
  by the corresponding box-product \eqref{eq:box-prod-def}
  \begin{align} \label{eq:calo-op-def}
    \mathcal{O} S_R = \Lambda_{r,a}^R (N) \cdot S_R
  \end{align}
  and differential operator $r \frac{\partial}{\partial p_r}$ acts
  in Schur basis in manifest way
  \begin{align}\label{eq:ddpr-op-schur-action}
    r \frac{\partial}{\partial p_r} S_R
    = (-1)^r \sum_{R' = R-\text{rim hook}} \frac{\sigma_r(R)}{\sigma_r(R')} S_{R'},
  \end{align}
  at least when $R$ has trivial $r$-core.
  Here $\sigma_r(R)$ is the $r$-signature of the diagram $R$.

  \bigskip

  With these definitions, one can check that a number
  of notable properties holds:
  \begin{itemize}
  \item
    Average of $K_\Delta$ with Schur function $S_R$ is equal to
    \begin{align}\label{eq:bilinear-superintegrability}
      \boxed{
      \left\langle K_\Delta S_R \right\rangle
      = (-1)^{\pi_{r,a,b}(\Delta)} S_{R/\pi_{r,a,b}(\Delta)} \{\delta_{k, r}\} \cdot \Lambda^R_{r,a}(N)
      }
    \end{align}
    Here $S_{R/Q}$ is the skew Schur polynomial, which we again evaluate at special
    point $p_k = \delta_{k,r}$\footnote{The appearance of skew Schur polynomials makes the story
    similar both to exotic sector of monomial MMs \cite{paper:BP-null-a-superintegrability}
    and to recently discovered large WLZZ family of MMs
    \cite{paper:MMMPZ-on-kp-integrable-skew-hurwitz-tau-functions-and-their-beta-deformations}.
    }. The permutation operation $\pi_{r,a,b}(\Delta)$ is a certain permutation
    on the space of partitions, that is somehow important to the story
    (it appears in several places, see below), and which we describe in detail
    in Section~\ref{sec:permutation-operation}.
    The $(-1)^{\pi_{r,a,b}(\Delta)}$ is the certain sign related to permutation
    $\pi_{r,a,b}$ which we also describe in Section~\ref{sec:permutation-operation}.
  \item
    As an elementary corollary of the previous property, the single-average
    of $K$-polynomial is trivial unless this polynomial corresponds to empty partition
    \begin{align} \label{eq:single-k-average}
      \left\langle K_{\Delta} \right\rangle
      \equiv \left\langle K_{\Delta} S_\emptyset \right\rangle
      = \delta_{\Delta,\emptyset}
    \end{align}
  \item
    The double-average of two $K$-polynomials $K_{\Delta_1}$ and $K_{\Delta_2}$
    is equally concise and manifest
    \begin{align} \label{eq:double-k-average}
      \boxed{
      \left\langle K_{\Delta_1} K_{\Delta_2} \right\rangle = \delta_{\Delta_1,\pi_{r,a,b}(\Delta_2)} \cdot 
      \sigma_r(\Delta_1) \sigma_r(\Delta_2) \cdot \Lambda_{r,a}^{\Delta_1}
      }
    \end{align}
    in case both $\Delta_1$ and $\Delta_2$ have trivial $r$-cores.
    The permutation operation $\pi_{r,a,b}(\Delta)$ is such that $\Lambda_{r,a}^{\Delta_1} = \Lambda_{r,a}^{\Delta_2}$ so it does not matter
    which one to use. In particular, when number of boxes is not equal,
    $|\Delta_1| \neq |\Delta_2|$, the bilinear $K$-average is always zero -- the feature that
    we originally used to calculate $K_\Delta$ polynomials recursively, before
    we understood the simple general formula \eqref{eq:k-delta-def}.
    
    In case only one of $r$-cores is non-trivial the average is zero.
    
    On the other hand, when both $r$-cores are non-trivial, there is also a non-trivial interaction
    structure, that even relaxes the selection rule $|\Delta_1| = |\Delta_2|$.
    For instance, for $r=3$ partitions $[2,2,1,1]$
    and $[3,2,2,1,1]$ both are their own non-trivial $r$-cores.
    At the same time, for $a=1$ $b=0$ we have
    \begin{align}
      \left\langle K_{[3,2,2,1,1]} K_{[2,2,1,1]} \right\rangle \neq 0
    \end{align}
    We present more examples of this non-trivial interaction in
    Section~\ref{sec:double-non-triv-cores}, but the general picture is, so far, missing.
    
  \end{itemize}
}

{\section{Permutation operation $\pi_{r,a,b}$}
  \label{sec:permutation-operation}

  The permutation operation $\pi_{r,a,b}$ is manifestly given by the following construction.

  For any partition $\Delta$ with trivial $r$-core, consider its
  $r$-quotients $\Delta_i$, $i=0\dots r-1$. $\pi_{r,a,b}$ rearranges $r$-quotients
  $\Delta_i$ according to the rule
  \begin{align} \label{eq:quotients-reshuffling-rule}
    \Delta_i \longrightarrow \Delta_{r-1 - i + a - 2 b \text{ mod } r}
  \end{align}
  and then partition $\Delta^{'} = \pi_{r,a,b}(\Delta)$ is reassembled from
  the shuffled parts.

  For instance, for $r=5$, $a=1$ $b=0$ then partition $[2,2,2,2,2]$
  has $5$-quotients: $(\emptyset,\emptyset,\emptyset,[1],[1])$.
  The reshuffling of quotients according to prescription
  \eqref{eq:quotients-reshuffling-rule} yields $(\emptyset,[1],\emptyset,\emptyset,[1])$
  while is $5$-quotient representation for partition $[4,2,2,1,1]$.
  Therefore, under $\pi_{5,1,0}$ we have
  \begin{align}
    [2,2,2,2,2] \longleftrightarrow [4,2,2,1,1]
  \end{align}

  In the Gaussian case $r=2$ the effect of $\pi_{r,a,b}$ operation is \textbf{not observed},
  since, for every $r$,$a$,$b$ one of $\Delta_i$'s always stays on its place,
  and so for $r=2$ does the only other.

  \bigskip
  
  \textbf{The sign} of the operation, $(-1)^{\pi_{r,a,b}(\Delta)}$,
  is calculated as follows. The overall sign is the product of the signs associated
  to elementary transpositions. For every $\Delta_i$ and $\Delta_{i'}$ that are being
  interchanged by $\pi_{r,a,b}$ they are either equal or different. Then
  \begin{align}
    \left\{
    \substack{
    \text{permutation} \\
    \text{contribution}
    }
    \right\} = 
    \left\{
    \begin{array}{l}
      1, \text{ if } \Delta_i = \Delta_{i'} \\
      (-1)^{i-i'}, \text{ if } \Delta_i \neq \Delta_{i'}
    \end{array}
    \right.
  \end{align}
}

{\section{Double-$K$ average in case of non-trivial cores} \label{sec:double-non-triv-cores}

  The formula \eqref{eq:k-delta-def} can be equally well applied when
  $\Delta$ has trivial or non-trivial $r$-core. When partition is its
  \textbf{o}wn $r$-\textbf{c}ore (denote it $\Delta_{oc}$),
  the corresponding Schur polynomial does not depend on $p_r$, and therefore
  $K$-polynomial is equal to Schur polynomial
  \begin{align}
    K_{\Delta_{oc}} = S_{\Delta_{oc}}
  \end{align}

  The structure of pair correlators of such partitions is much less obvious
  than simple formula \eqref{eq:double-k-average}: here we list some
  more-or-less astonishing examples:

  \begin{itemize}
  \item Some polynomials are ``vanishing'' vectors -- orthogonal to every partition
  with same number of boxes, including itself. For instance, for $r=3,\ a=1,\ b=0$:
  \begin{align}
    \left\langle K_{[2,2,1,1]} K_{[2,2,1,1]} \right\rangle = 0
    \ \ \left\langle K_{[2,2,1,1]} K_{R} \right\rangle = 0, \text{ for } |R|=6
  \end{align}
  \item
    At the same time, the average between partitions with different $r$-cores
  and different number of boxes is non-vanishing
  \begin{align}
    \left\langle K_{[2,2,1,1]} K_{[3,2,2,1,1]} \right\rangle
    = \frac{1}{243} N^2 (N+1)(N-2)(N-3)
  \end{align}
  Note that the $N$-dependent factor is equal to
  $\Lambda_{3,0}^{[2,2,1,1]}(N)=\Lambda_{3,0}^{[3,2,2,1,1]}(N)$,
  that is, it looks like
  \begin{align}
    \text{Non-trivial are the pair correlators
    between $K$-polynomials that have \textit{coincident} $\Lambda$-factors.}
  \end{align}
  Whether this is actually true or not, remains to be seen in a separate
  thorough study.

  \item
    Furthermore, the non-vanishing correlators get even more complicated.
  For instance, both quadratic (i.e. same $\Delta$)
  \begin{align}
    \left\langle K_{[7,2]} K_{[7,2]} \right\rangle
    = \frac{(-1)}{9}\left(N^2 + 10 N + 33\right)\Lambda_{3,0}^{[7,2]}(N) \\ \notag
    \left\langle K_{[4,2,2,1]} K_{[4,2,2,1]} \right\rangle
    = \frac{(-1)}{9} (N-3)(N-2) \Lambda_{3,0}^{[4,2,2,1]}(N)
  \end{align}
  and bilinear correlators
  \begin{align}
    \left\langle K_{[4,2,2,1]} K_{[4,2,1,1,1]} \right\rangle
    = \frac{1}{9}\left(N^2 - 5 N + 15\right) \Lambda_{3,0}^{[4,2,2,1]}(N)
  \end{align}
  can have extra, often non-factorizable, factors
  (in addition to being divisible by the usual $\Lambda$-factor).

  \end{itemize}
  
  It remains to be seen, whether these extra (non-factorizable) factors
  can be amended by some clever redefinition of $K$-polynomials in case
  of non-trivial cores; or, perhaps, some more general clever formula can be
  invented that will take into account these more compilcated cases \textit{as is}.
}

{\section{Triple-$K$ averages} \label{sec:triple-k-averages}
  The single- and double-$K$ averages are reminiscent to the averages in
  conformal field theory, where, for the primary operators one has
  \begin{align}
    \left\langle \mathcal{O}(x) \right\rangle \sim & \ \delta_{\Delta,0} \\ \notag
    \left\langle \mathcal{O}_1(x) \mathcal{O}_2(y) \right\rangle
    \sim & \ \frac{\delta_{\Delta_1,\Delta_2}}{(x-y)^{\Delta_1 + \Delta_2}},
  \end{align}
  where, $\Lambda_{r,a}^{\Delta_{1,2}}(N)$ in \eqref{eq:double-k-average}
  can be, perhaps, thought of as ``discrete'' analog of $(x-y)^{-\Delta_1-\Delta_2}$.

  In this logic, the simple form of the three-point average in conformal field theory
  \begin{align}
    \left\langle \mathcal{O}_1(x) \mathcal{O}_2(y) \mathcal{O}_3(z) \right\rangle
    = \frac{C_{\Delta_1,\Delta_2,\Delta_3}}{
      (x-y)^{\Delta_1 + \Delta_2 - \Delta_3}
      (y-z)^{\Delta_2 + \Delta_3 - \Delta_1}
      (z-x)^{\Delta_3 + \Delta_1 - \Delta_2}}
  \end{align}
  should imply, on our matrix model side,
  comparably simple fully-factorized
  triple-$K$ average, where $N$-dependence is made from peculiar
  combinations of $\Lambda_{r,a}^{\Delta_{1,2,3}}(N)$-factors.

  This naive hope, is, however, overoptimistic. While for some
  small digrams the average, indeed, is factorizable and simple.
  For instance, for $r=3, a=1, b=0$
  \begin{align}
    \left\langle K_{[3]} K_{[3]} K_{[3]} \right \rangle
    = -\frac{2}{27} N (N+1)(N+2).
  \end{align}

  For other diagrams the average stops being factorizable
  \begin{align}
    \left\langle K_{[6]} K_{[6]} K_{[5,1]} \right \rangle
    = -\frac{2}{729} N (N + 1) (N + 4) (N + 3) (N^2  + 16 N + 57)
  \end{align}

  The non-factorization, however, seems at the moment to be mild: in the examples we analyzed
  at most quadratic non-factorized polynomial was observed.
  Therefore, it can yet turn out that three-point $K$-average is
  always a sum of at most two fully-factorized expressions.
  For instance, with the above example the plausible ``split'' could look like
  \begin{align}
    \left\langle K_{[6]} K_{[6]} K_{[5,1]} \right \rangle
    = -\frac{2}{729} N^2 (N + 1) (N + 4) (N + 3)^2
    + \frac{2 \cdot 19}{729} (N-3) N (N + 1) (N + 4) (N + 3),
  \end{align}
  where one now needs to explain the origin of the two summands.

  \bigskip
  
  Further intensive studies are needed to discern between several alternatives,
  which are equally probable at the moment:
  \begin{itemize}
  \item the non-factorizability of triple $K$-polynomial average is, indeed,
    at most quadratic. Some hidden structure (perhaps, an analog of KZ-equation or similar)
    is controlling this simplification;
  \item the proper matrix model analog of primary operators are not just $K_\Delta$
    polynomials with trivial-core $\Delta$, but $K_\Delta$'s with some
    additional condition/requirement. The triple averages of such, ``truly primary'',
    $K_\Delta$'s are factorizable, while averages of ``descendent'' $K_\Delta$,
    in general, do not factorize;
  \item the triple $K$-polynomial averages are fully non-factorizable and generic,
    and no hidden structure exists.
  \end{itemize}
}

{\section{Towards proofs}
  \label{sec:proof-attempts}

  The experimentally observed bilinear superintegrability formulas
  \eqref{eq:bilinear-superintegrability},
  \eqref{eq:single-k-average} and \eqref{eq:double-k-average} are crisp and
  concise. One may, therefore, be tempted to think that their \textit{proof}
  is equally crisp and simple, and follows from ready generalizations of
  certain MM/representation-theoretic constructions to the monomial case.

  At least at the moment this does not seem to be the case: several attempts
  (listed below) to find such auxiliary generalized structures that would help
  in the proof, fail. This, of course, makes the bilinear superintegrability formulas
  \eqref{eq:bilinear-superintegrability},
  \eqref{eq:single-k-average} and \eqref{eq:double-k-average}
  all the more interesting and valuable: true examples of emergent structure,
  which cannot be naively reduced to/explained by more fundamental observations.
  
  {\subsection{The first encouraging successes}
    \begin{itemize}
    \item
    The single K-average \eqref{eq:single-k-average} is, quite naturally, simpler
    than bilinear \eqref{eq:bilinear-superintegrability} and \eqref{eq:double-k-average},
    so one may hope to prove it first.

    And indeed
    \begin{align} \label{eq:single-k-delta-proof-1}
      \left\langle K_\Delta \right\rangle =
      \left\langle \mathcal{O}^{-1}
      \exp\left((- r) \frac{\partial}{\partial p_r} \right)
      \mathcal{O} S_\Delta \right\rangle
      \mathop{=}_{\eqref{eq:k-delta-def}-\eqref{eq:ddpr-op-schur-action},
        \eqref{eq:monomial-schur-average}}
      \sum_{\nabla \in_r \Delta} \frac{\xi_\Delta}{\xi_\nabla}
      (-1)^{r + \sigma(\Delta-\nabla)} Q(\Delta-\nabla)
      S_\nabla\{\delta_{k,r}\} \frac{1}{r^{|\nabla|/r}} \xi_\nabla,
    \end{align}
    where $\in_r$ means summation over diagrams obtained from $\Delta$ by removing some
    $r$-rim-hooks, and $Q(\Delta-\nabla)$ is the number of ways to obtain $\nabla$
    from $\Delta$ by doing so. Now, continuing
    \begin{align} \label{eq:single-k-delta-proof-2}
      \eqref{eq:single-k-delta-proof-1} = \sum_{\nabla \in_r \Delta}
      (-1)^{r + \sigma(\Delta-\nabla)} Q(\Delta-\nabla)
      \frac{1}{r^{|\nabla|/r}} S_\nabla\{\delta_{k,r}\}
      \xi_\Delta
      = \xi_\Delta \left(
      \exp\left(-\frac{\partial}{\partial p_r} \right) S_\Delta(p)
      \right)_{p_k=\delta_{k,r}} = \delta_{\Delta, \emptyset}
    \end{align}

  \item Similarly, the implication
    $\eqref{eq:bilinear-superintegrability} \rightarrow \eqref{eq:double-k-average}$
    is easy to prove. Indeed, expanding the definition
    \begin{align}
      \left\langle K_\Delta K_{\Delta^{'}} \right\rangle
      = & \ \sum_{R \in_r \Delta^{'}} \left\langle K_\Delta S_R \right\rangle
      (-1)^{r + \sigma(\Delta^{'} - R)} Q(\Delta^{'} - R)
      \frac{\xi_{\Delta^{'}}}{\xi_R}
      \\ \notag
      = & \ \xi_{\Delta^{'}} \left (
      (-1)^{\pi(\Delta)} S_{R/\pi(\Delta)} \left\{\delta_{k,r}\right\}
      (-1)^{r + \sigma(\Delta^{'} - R)} Q(\Delta^{'} - R)
      \right ),
    \end{align}
    where the sum in brackets is independent of $N$, and with more combinatorial
    massaging of the skew Schur functions analogous to
    \eqref{eq:single-k-delta-proof-2} we prove the sign and a selection rule.

    \end{itemize}
    
    \bigskip \bigskip

    \noindent Writing down the bilinear average in a similar manner
    \begin{align}
      \left\langle K_\Delta S_R \right\rangle =
      \xi_R \left(\sum_{\substack{\nabla \in_r \Delta \\ P}}
      (-1)^{\sigma(\Delta-\nabla)} Q(\Delta-\nabla)
      \cdot N_{\nabla R}^P \cdot \frac{\xi_\Delta \xi_P}{\xi_\nabla \xi_R} \cdot
      \frac{1}{r^{|P|/r}} S_P \{\delta_{k,r}\} \right),
    \end{align}
    where $N_{\nabla R}^P$ are the Littlewood-Richardson coefficients, we see that
    the goal is, firstly, to prove that the sum in brackets is $N$-independent
    and, secondly, that the peculiar permutation operator $\pi_{r,a,b}$ emerges.
    How to do this, however, at the moment is not at all obvious:
    for illustration we present here a couple of proof ideas that fail
    (i.e. the emergent structure \eqref{eq:bilinear-superintegrability}
    is not decomposable into/explained by these, simpler, putative sub-structures).
  }

  {\subsection{No Cauchy-like summation}
    There is the following formula for the summation of the skew Schur functions
    \cite{book:M-symmetric-functions-and-hall-polynomials}
    \begin{align}
      \sum_{\Delta \in R} S_{R/\Delta} (p) S_{\Delta}(g) = S_R(p+g)
    \end{align}
    which simplifies the r.h.s of \eqref{eq:bilinear-superintegrability},
    provided one rewrites the permutation $\pi_{r,a,b}$ and the sign to the left hand side.

    Then the hope would be, that the corresponding l.h.s sum
    \begin{align}
      P_{r,a,b}(R) := \sum_{\Delta \in R} K_{\pi_{r,a,b}(\Delta)} (p)
      \cdot (-1)^{\pi_{r,a,b}(\Delta)}
      \cdot
      S_{\Delta}(g)
    \end{align}
    actually evaluates to something nice and concise.

    This, however, turns out not to be the case as first few examples
    \begin{align}
      P_{3,1,0}([3]) = & \ 
      \frac{1}{54} \left(-3 p_1^3 - N^2 - N + 3 p_3 \right) g_1^3
      - \frac{1}{18} g_2 \left(3 p_1^3 + N^2 + N - 3 p_3 \right) g_1
      - \frac{1}{27} g_3 N^2 - \frac{1}{27} g_3 N + 1
      \\ \notag
      + & \ \frac{1}{54} \left (-6 p_1^3 + 6 p_3 \right ) g_3
      \\ \notag
      P_{3,1,0}([2,1]) = & \ 
      \frac{1}{54} \left(-3 p_1^3 + 2 N^2 - 9 p_1 p_2 + 2 N - 6 p_3 \right) g_1^3
      + 1 + \frac{1}{54} \left(3 p_1^3 - 2 N^2 + 9 p_1 p_2 - 2 N + 6 p_3 \right) g_3
      \\ \notag
      P_{3,1,0}([1,1,1]) = & \ 
      \frac{1}{108} \left(3 p_1^3-2 N^2-9 p_1 p_2 + 4 N + 6 p_3 \right) g_1^3
      +\frac{1}{18} \left(-\frac{3}{2} p_1^3 + N^2
      + \frac{9}{2} p_1 p_2 - 2 N - 3 p_3 \right) g_2 g_1
      \\ \notag
      + & \ \frac{1}{18} g_3 p_1^3 - \frac{1}{27} g_3 N^2
      - \frac{1}{6} g_3 p_1 p_2 + \frac{2}{27} g_3 N + \frac{1}{9} g_3 p_3 + 1
    \end{align}
    reveal no apparent structure.
  }

  {\subsection{No Littlewood-Richardson structure}
    Another approach would be to go via the orbifoldization construction
    of \cite{paper:CHPS-orbifolds-and-exact-solutions}, eqn.(4.33). From that point of view
    the single Schur average turns out to be the product over the $r$-quotients.
    \begin{align}
      \left\langle
      S_R
      \right\rangle_N \sim \prod_{i=0}^{r-1} \left\langle
      S_{R^{(i)}}
      \right\rangle_{n_i, u_i},
    \end{align}
    where $R^{(i)}$ are the $r$-quotients of $R$ and the correlators on the l.h.s.
    are evaluated in the simpler logarithmic model.

    For the proof along these lines to go through two crucial things need to happen.
    First, the expression for $K_\Delta$ polynomial should be reasonably simple
    in this language of $r$-quotients.

    Secondly, Schur polynomial multiplication (i.e. Littlewood-Richardson coefficients),
    at least in the trivial $r$-core case, should be ``consistent'' with $r$-quotient
    language: the result should be expressed through individual $r$-quotients in
    a reasonable way.

    \bigskip

    The first crucial thing is, indeed, true. On one hand, the $\mathcal{O}$-operator
    eigenvalue $\xi_R$ is (analogously to orbifoldization construction)
    expressed through Schur functions for the respective $r$-quotients.
    On the other hand, the shift operator $\exp((- r) \partial/\partial r)$
    acts by removing $r$-rim-hooks in all possible ways, which in the language of
    $r$-quotients is nothing but removing all boxes in all possible ways
    (with suitable signs).

    \bigskip

    However, the second crucial thing seems not to be the case. For instance,
    multiplying two partitions $[3]$ and $[2,1]$, which, in the language of $r=3$-quotients
    are equal to $([1], \emptyset, \emptyset)$ and $(\emptyset, [1], \emptyset)$
    one gets
    \begin{align}
      [3] \otimes [2,1] = & \ [5,1] + [4,2] + [4,1,1] + [3,2,1] \\ \notag
      ([1], \emptyset, \emptyset) \otimes (\emptyset, [1], \emptyset)
      = & \ (\emptyset, [2], \emptyset) + (\emptyset, \emptyset, \emptyset)_{[4,2]}
      + (\emptyset, \emptyset, [2]) + ([1], \emptyset, [1]),
    \end{align}
    i.e. (even omitting the appearance of non-trivial $r$-core diagrams, which vanish
    later in the correlator) boxes are merged and shuffled in obscure ways.
    This gets even more complicated for bigger partitions.

    \bigskip
    
    Other plausible, but equally barren, proof strategies (for instance
    the study of interplay between $\mathcal{O}$-operator and the Littlewood-Richardson
    coefficients) are possible but we don't list them here.
    In any case, desired is not the technical proof, but rather the
    \textit{conceptual explanation} of why the bilinear superintegrability
    formula \eqref{eq:bilinear-superintegrability} is true.
  }
  
}

{\section{Conclusion}\label{sec:conclusion}
  In this paper we studied, to what extent the recently proposed bilinear
  superintegrability \cite{paper:MM-bilinear-character-correlators-in-superintegrable-theory}
  persists in the case of matrix models in pure phase
  \cite{paper:BP-null-a-superintegrability,paper:CHPS-orbifolds-and-exact-solutions}.

  We found, that in the case of trivial $r$-cores, it generalizes simply and naturally,
  according to formula \eqref{eq:bilinear-superintegrability}.
  Moreover, the associated $K_\Delta$ polynomials are obtained with help
  of triangular change of variables \eqref{eq:k-delta-def},
  where the central ingredient (the $\mathcal{O}$-operator) is, as well,
  a natural monomial generalization with respect to the Gaussian case.

  The key prominent feature of bilinear superintegrability in the monomial case
  is the appearance of non-trivial permutation operation $\pi_{r,a,b}$
  (see Section~\ref{sec:permutation-operation}), which trivializes in Gaussian case
  but generally is expressed in the language of Young diagram $r$-quotients.
  This non-trivial permutation operation, arguably, is the reason why
  the bilinear superintegrability formula for monomial
  non-(q,t)-deformed models was not found during the earlier attempts
  \cite{paper:CHPS-orbifolds-and-exact-solutions,
    paper:MMS-a-direct-proof-of-agt-at-beta-1,
    AFLT-on-combinatorial-expansion-of-the-conformal-blocks,
    paper:MMS-proving-agt-as-hs-duality}.

  Finally, the explicit and simple form of bilinear superintegrability
  in the language of $K_\Delta$ polynomials allowed us,
  in Section~\ref{sec:triple-k-averages}
  to pose some questions about general analogy between matrix models
  and conformal field theories,
  beyond the well-known AGT conjecture,
  and in the spirit of recent attempt to generalize Nekrasov calculus
  beyond AGT
  \cite{paper:MM-superintegrability-as-the-hidden-origin-of-the-nekrasov-calculus}.
  We performed just a few naive comparison attempts
  and they show that this matrix model \textit{conformaliztion} program
  is not straightforward and immediate, yet, it is not immediately ruled out.
  We hope to study the situation in detail in the future.

  \bigskip

  Few immediate concrete questions seem natural in the context of the present paper:
  \begin{itemize}
  \item What is the manifest expression of the operator $\mathcal{O}$
    in terms of time variables $p_k$? Naive symbolic experiments show that
    $\mathcal{O}(p)$ likely is of infinite degree w.r.t derivatives in $p_k$.
  \item How does the story generalize to exotic sector? Both in the ``strong'' sense
    of \cite{paper:BP-null-a-superintegrability},
    where the role of normalization constant is played not by partition function
    $Z = \left\langle 1\right\rangle$, and in the ``weak'' sense
    of Section~\ref{sec:double-non-triv-cores} where non-vanishing core partitions
    interact on the trivial core ``background'' of the basis Schur correlators.
    Is there any similarity at all between descriptions of these ``strong'' and
    ``weak'' exotic sectors?
  \item What is the proper $q-$ and $\beta-$ deformation of the associated $K$-polynomials
    and what shape does their bilinear superintegrability take?
    How does it relate to the long-known formula for double-Schur/Jack correlator
    in these models (which does not seem to have $q\rightarrow1, \ \beta\rightarrow1$ limit)?
  \item Is the appearance of at most quadratic non-factorizable polynomials a general
    feature of multiple-$K$ averages in monomial matrix models, or is it just an
    artifact of the partitions with small number of boxes?
  \end{itemize}

  \bigskip

  All these intriguing questions will hopefull be studied in the future.
  
}

{\section*{Acknowledgments}
  We are grateful to A.Mironov, A.Morozov and Pei-Ming Ho for stimulating discussions.
  Our work is partly supported by the grant of the Foundation for the Advancement of Theoretical Physics “BASIS”, by the joint RFBR grant 21-51-46010-ST-a,
  by the joint RFBR-MOST grant 21-52-52004 MNT\_a.
  Chuan-Tsung Chan is supported in part by the NSCT of Taiwan through the grant number
  110-2923-M-002-016-MY3.
}

\bibliographystyle{mpg}
\bibliography{references_shadow-phase-mon-matr}

\end{document}